# The Sun's Open–Closed Flux Boundary and the Origin of the Slow Solar Wind

Chloe P. Wilkins[1], David I. Pontin[1], Anthony R. Yeates[2], Spiro K. Antiochos[3], Hannah Schunker[1], and Bishnu Lamichhane[1]
[1] School of Information and Physical Sciences, University of Newcastle, University Drive, Callaghan, NSW 2308, Australia
[2] Department of Mathematical Sciences, Durham University, Durham, DH1 3LE, UK
[3] Department of Climate and Space Sciences and Engineering, University of Michigan, Ann Arbor, MI 48109, USA
*Received 2025 March 7; revised 2025 April 13; accepted 2025 April 14; published 2025 May 23*

## Abstract

The Sun's open–closed flux boundary (OCB) separates closed and open magnetic field lines, and is the site for interchange magnetic reconnection processes thought to be linked to the origin of the slow solar wind (SSW). We analyze the global magnetic field structure and OCB from 2010 December to 2019 December using three coronal magnetic field models: a potential-field source-surface (PFSS) model, a static equilibrium magnetofrictional model, and a time-dependent magnetofrictional model. We analyze the model and cycle dependence of the OCB length on the photosphere, as well as the magnetic flux in the vicinity of the OCB. Near solar maximum, the coronal magnetic field for each model consists predominantly of long, narrow coronal holes, and nearly all the open flux lies within 1 supergranule diameter (25 Mm) of the OCB. By comparing to interplanetary scintillation measurements of SSW speeds, we argue that the fraction of open flux within this 25 Mm band is a good predictor of the amount of SSW in the heliosphere. Importantly, despite its simplicity, we show that the PFSS model estimates this fraction as well as the time-dependent model. We discuss the implications of our results for understanding SSW origins and interchange reconnection at the OCB.

*Unified Astronomy Thesaurus concepts:* Slow solar wind (1873); Solar corona (1483); Solar coronal holes (1484); Solar magnetic fields (1503)

## 1. Introduction

The source region and acceleration of the slow solar wind (SSW) remain fundamental questions in solar physics, with significant implications for understanding space weather and its impact on Earth. The SSW is distinguishable from the fast solar wind by variabilities in plasma composition and properties such as elemental abundances, ion charge-state ratios, and wind speeds (typically $<500$ km s$^{-1}$ for the SSW; e.g., S. K. Antiochos et al. 2011; L. Abbo et al. 2016). While the fast solar wind is thought to originate from the centers of large coronal holes (CHs), the origin of the SSW is less certain.

The Sun's global coronal magnetic field can be classified in terms of "closed" field lines (where both ends are rooted on the photosphere) and "open" field lines (where only one end is rooted on the photosphere and the field line extends indefinitely into the heliosphere). Observations indicate that the SSW and closed corona share similar ratios of elements with low-to-high first ionization potentials (e.g., T. H. Zurbuchen & R. von Steiger 2006; S. K. Antiochos et al. 2011), suggesting that the SSW plasma originates within closed magnetic flux, but is then accelerated along open flux. Owing to the high conductivity of the coronal plasma, the only way for the plasma to transfer from the magnetically closed to the magnetically open region is for magnetic reconnection to reconfigure the magnetic field. *Interchange* magnetic reconnection is the specific name given to reconnection between open and closed flux (N. U. Crooker et al. 2002), and is the leading hypothesis for explaining the origin of the SSW (e.g., S. K. Antiochos et al. 2011; D. I. Pontin & P. F. Wyper 2015; Y. M. Wang 2024). As such, a characterization of the interface between open and closed flux—that we term the *open–closed boundary*, or *OCB* —is crucial for exploring the link between interchange reconnection and the origin of the SSW.

The Sun's magnetic field evolves over the 11 yr solar cycle. During solar minima, the large-scale structure of the field is dominated by a dipole component, meaning that large, long-lived CHs are present in the polar regions while coronal loops are concentrated near the equator. Moving toward solar maximum, the polar CHs shrink and eventually disappear, while smaller, shorter-lived CHs appear at lower latitudes (M. Miralles et al. 2004). The increase in active regions on the photosphere at solar maximum leads to a more complex field structure in the overlying atmosphere. As a result, higher-order spherical harmonics—representing more localized, smaller-scale magnetic features—become more significant in the description of the Sun's large-scale field. Furthermore, it is well known that the total quasi-steady open flux measured in the heliosphere is considerably larger during maximum than during minimum (C. N. Arge et al. 2024). To characterize the Sun's OCB, it is therefore important to explore the magnetic field structure across different phases of the solar cycle. In this study, we focus on the period 2010 December to 2019 December within Solar Cycle 24.

The structure of the OCB is naturally influenced by the choice of coronal magnetic field model. Different approaches exist for constructing a model coronal field (D. H. Mackay & A. R. Yeates 2012; T. Wiegelmann et al. 2017), with the simplest and most popular model being a so-called potential-field source-surface (PFSS) model. This model has the advantage of being cheap to compute and having a unique solution for the given boundary data and source-surface height. However, it neglects the free magnetic energy in the corona and the effects of the plasma, and is known to have issues, for example in matching the observed open flux measured in situ (J. A. Linker et al. 2017). We therefore explore three distinct models: a PFSS model, a nonpotential model that incorporates a solar wind velocity profile, and a time-dependent model that includes both nonpotentiality and surface flux transport. We demonstrate the effect of these assumptions on the length of the OCB on the photosphere and the surrounding magnetic flux.

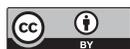

Previous studies (e.g., N. A. Schwadron et al. 2005; S. K. Antiochos et al. 2011; Y. M. Wang 2024) suggest that SSW streams are formed by interchange reconnection between the outermost closed flux and the adjacent open flux at the *edges* of CHs. Given the strong correlation between ratios of the abundances of elements in the closed corona and the SSW, these findings suggest that the closed flux near CH boundaries likely influences the composition of interchange-released plasma, which is subsequently transported into the heliosphere along open field lines adjacent to the boundary. We are therefore interested not only in the OCB itself, but also in the open and closed magnetic flux near the OCB.

In this study, we characterize the dependence of the Sun's OCB on both the coronal magnetic field model and the solar cycle phase. Section 2 outlines the three coronal magnetic field models analyzed, and details our method for defining and identifying the OCB. In Section 3, we discuss variations in the size and flux contributions of midlatitude CHs and their effect on the overall length of the OCB on the photosphere. In Section 4, we examine the magnetic flux near the OCB in both the closed and open magnetic field regions, comparing our findings with sunspot numbers and interplanetary scintillation (IPS) measurements of solar wind speeds. We finish in Section 5 with our conclusions.

## 2. Coronal Magnetic Field Models and Methodology

To characterize the Sun's OCB and its dependence on different coronal magnetic field models, we explore three distinct methods for modeling the Sun's global field. These methods vary in complexity and in their underlying assumptions, allowing us to assess how factors such as nonpotentiality and surface flux transport influence the OCB structure over the solar cycle.

The three models chosen for this study are a time-dependent magnetofrictional (TDMF) model, a static equilibrium magnetofrictional (SEMF) model, and a PFSS model. The latter is the simplest and most commonly used, but does not account for free magnetic energy in the corona. On the other hand, both magnetofrictional models allow for nonzero currents and maintain a balance between the Lorentz force and an outflowing wind. The TDMF model further incorporates the time history of the coronal evolution using a photospheric boundary driven by surface flux transport. The three models are detailed below.

### 2.1. Time-dependent Magnetofrictional Model

The TDMF model we use is the Durham Magneto-Frictional Code (DuMFriC), a magnetofrictional solution coupled with surface flux transport in which the large-scale/mean magnetic field $\boldsymbol{B}$ is evolved over time $t$ according to Faraday's law:

$$\frac{\partial \boldsymbol{B}}{\partial t} = -\nabla \times \boldsymbol{E}, \quad (1)$$

where the electric field $\boldsymbol{E}$ is given by

$$\boldsymbol{E} = -\boldsymbol{v} \times \boldsymbol{B} + \boldsymbol{N}. \quad (2)$$

Here, the effect of small-scale fluctuations in the magnetic field is represented by the small, nonideal hyperdiffusive term $\boldsymbol{N}$ (see A. R. Yeates 2024 for details). The velocity $\boldsymbol{v}$ is given by

$$\boldsymbol{v} = \frac{(\nabla \times \boldsymbol{B}) \times \boldsymbol{B}}{\nu |\boldsymbol{B}|^2} + v_\mathrm{w}\left(\frac{r}{R_\mathrm{outer}}\right)^{11.5} \boldsymbol{e}_r. \quad (3)$$

The first term here involving the Lorentz force allows for a magnetofrictional relaxation toward a force-free equilibrium, and the rate of this relaxation is controlled by the friction coefficient $\nu$. The second term models the effects of the solar wind in the outer corona with a radial outflow distribution, balancing the Lorentz force term. The TDMF simulations considered in this paper use $\nu = \nu_0 (r \cos \lambda)^{-2}$, where $\nu_0 = 2.8 \times 10^5$ s for radius $r$ and latitude $\lambda$; an outflow speed coefficient $v_\mathrm{w} = 100$ km s$^{-1}$; and a fixed outer radius $R_\mathrm{outer} = 2.5 R_\odot$, where $R_\odot$ is the radius of the Sun. The simulations were initialized with a PFSS extrapolation of smoothed radial magnetic field data from the `hmi.synoptic_mr_polfil_720s` series for Carrington rotation CR2097, taken by the Helioseismic and Magnetic Imager (HMI; P. H. Scherrer et al. 2012) from the Solar Dynamics Observatory.[4] The surface flux transport scheme in the TDMF model includes the emergence of active regions with an associated nonuniform twist related to $B_z$ and $J_z$ (determined from HMI data), where $\boldsymbol{J} = \nabla \times \boldsymbol{B}$ is the electric current density. Full parameter values and model details for these simulations are provided in A. R. Yeates (2024). Specifically, we use the simulation run "TOb5" from that paper.

### 2.2. Static Equilibrium Magnetofrictional Model

The SEMF model we consider in this paper is detailed in O. E. K. Rice & A. R. Yeates (2021). The basis for this model is a static equilibrium solution to Equation (1) using the ideal electric field

$$\boldsymbol{E} = -\boldsymbol{v} \times \boldsymbol{B}, \quad (4)$$

where $\boldsymbol{v}$ is similarly given by Equation (3). We adopt a constant friction coefficient $\nu = 5 \times 10^{-17}$ s cm$^{-2}$ to match previous SEMF simulations (see O. E. K. Rice & A. R. Yeates 2021).

The SEMF model solves for the magnetofrictional equilibrium given by $\boldsymbol{v} \times \boldsymbol{B} = \boldsymbol{0}$, subject to an imposed (time-independent) distribution of $B_r$ on the solar surface. Unlike the TDMF model, the SEMF model does not take into account the effects of low-coronal nonideal terms present in the electric field. Furthermore, while the TDMF model retains some "memory" of the field topology over time, the SEMF model produces an independent extrapolation at each time that is not causally related to the field at adjacent times.

### 2.3. Potential-field Source-surface Model

Finally, we consider a PFSS model given by the `pfsspy` software package.[5] Unlike the nonpotential models described above, the PFSS model neglects electric currents within the domain, which reduces Maxwell's equations of electromagnetism to

$$\nabla \cdot \boldsymbol{B} = 0 \text{ and } \nabla \times \boldsymbol{B} = \boldsymbol{0}. \quad (5)$$

---

[4] The data were obtained from the Joint Science Operations Center (JSOC) at http://jsoc.stanford.edu/.
[5] See https://pfsspy.readthedocs.io/en/stable/index.html for details.

The only free parameter for a PFSS solution is the height of the source surface, which we take to be $2.5R_\odot$. The main difference in the extrapolated magnetic field between the SEMF and PFSS models is near to the outer boundary, with the SEMF model retaining the distending effect of a solar wind outflow (O. E. K. Rice & A. R. Yeates 2021).

### 2.4. Boundary Conditions

For all three models, the magnetic field $\boldsymbol{B}$ is computed in a 3D spherical shell $R_\odot \leqslant r \leqslant 2.5R_\odot$. We confine ourselves to analyzing the field structure based on magnetogram resolutions of (180, 360) points equally spaced in $(\cos\theta, \phi)$, where $\theta$ and $\phi$ represent the colatitude and longitude, respectively. We solve for the magnetic field at 61 points equally spaced in $\log(r)$.

In the following sections, we describe the coronal magnetic field structure in the three models for *identical* photospheric flux distributions. The TDMF model is run for the period of 2010 June 12 to 2019 December 31. We exclude the first six months of this period while the nonpotential stresses build up in the volume, and then we select 124 maps every 27 days between 2010 December 18 and 2019 December 25, plus an additional map at 2019 December 31. For each snapshot, we extract the radial photospheric magnetic field distribution $B_r$ and use it as the photospheric boundary condition for the two static extrapolation models (PFSS and SEMF).

The outer boundary conditions vary between the three models. For the PFSS model, a source-surface boundary is imposed at $2.5R_\odot$, and the conditions

$$B_\theta \mid_{r=2.5R_\odot} = 0 \text{ and } B_\phi \mid_{r=2.5R_\odot} = 0 \quad (6)$$

are enforced to mimic the effects of the radial solar wind outflow beyond $2.5R_\odot$. This condition also ensures that all closed field lines lie within $r < 2.5R_\odot$.

For the SEMF model, a numerical source-surface boundary is similarly imposed at $2.5R_\odot$, however the actual maximum height of the closed field lines is determined by the radial outflow profile in Equation (3) rather than the precise height of this outer boundary.

For the TDMF model, the outer boundary is also defined at $2.5R_\odot$. Tangential motions driven by the magnetofriction term in Equation (3) can introduce artificial magnetic energy flux across the boundaries of the domain. To prevent this, the tangential components of the current density are set to zero at both the inner ($r = R_\odot$) and outer boundaries. In other words, the condition $\boldsymbol{J} \times \boldsymbol{e}_r = \boldsymbol{0}$ is imposed. On the lower boundary, an additional electric field is imposed to implement the surface flux transport scheme (see A. R. Yeates 2024 for details). There is also a zero-gradient condition imposed for the computation of the hyperdiffusive term $N$ in Equation (2) on both the inner and outer boundaries, detailed in A. R. Yeates (2024).

### 2.5. Effect of Boundary Conditions on the Magnetic Field Solution

The model assumptions and boundary conditions detailed above impact the magnetic field solution, including the shape and height of closed magnetic field lines and the similarities between the solutions and observations. This is illustrated in the top panel of Figure 1, which shows an example of the magnetic field extrapolation for the (left) PFSS model, (middle) SEMF model, and (right) TDMF model for the first simulation date in the series (i.e., 2010 December 18).

For PFSS models, the height of the closed field lines can be increased (decreased) by raising (lowering) the height of the outer source-surface boundary. As such, magnetic field extrapolations using PFSS models often depend heavily on the imposed height of the fixed source surface. Furthermore, streamers (i.e., closed field lines that separate regions of opposite magnetic polarity) in PFSS models extend right up to the source surface, where they often exhibit sharp bends or abrupt changes in curvature as the field lines transition from closed to open. This behavior is illustrated in the top-left panel of Figure 1, where a subset of magnetic field lines for the PFSS solution have been plotted. These sharp bends are an artifact of the assumption of a rigid source surface, which constrains the field lines to become radial at that height and leads to an artificial "kinking" of the closed field lines near the boundary.

However, coronal observations suggest that the coronal field does not become radial at a fixed height across all latitudes and throughout the solar cycle (B. Boe et al. 2020). One method for overcoming this is to incorporate a solar wind outflow velocity (as per the SEMF and TDMF models) as this allows for the closed field lines to extend to different altitudes depending on the strength of the local magnetic field. This is evident when comparing the top-left and top-middle panels of Figure 1, which show a subset of magnetic field lines for the PFSS and SEMF models initialized with the same lower boundary data. As suggested by O. E. K. Rice & A. R. Yeates (2021), it is evident in this figure that the addition of a sufficiently large outflow velocity prevents the closed streamers from extending up to the source surface, but instead has the field becoming radial at lower altitudes compared to the PFSS model. Although a numerical source-surface boundary is similarly imposed for the SEMF model, O. E. K. Rice & A. R. Yeates (2021) show that the height of this boundary does not significantly influence the magnetic field extrapolations when the outflow velocity is present (provided that the source surface is sufficiently high and the outflow velocity sufficiently large).

As with the SEMF model, the plasma flow in the TDMF model (top-right panel of Figure 1) is close to radial at the outer boundary (see Equation (3)) so that the magnetic field there is also close to radial. Again, and unlike the PFSS model, streamers can close below this outer boundary. Furthermore, since the source-surface condition is not imposed directly in the TDMF model, this also allows erupting magnetic structures to escape, temporarily enhancing the horizontal magnetic field before it relaxes back to near radial.

Evidently, boundary conditions significantly affect the structure of the extrapolated magnetic field. While PFSS models are computationally efficient, the rigid-source-surface assumption can result in abrupt transitions in the field-line geometry and limitations in modeling streamer heights. By contrast, the SEMF and TDMF models provide more realistic representations of coronal streamer shapes, with the TDMF model also accounting for dynamic processes like erupting magnetic structures. Compared to full MHD simulations, the SEMF and TDMF models are less computationally expensive and require only line-of-sight magnetogram data, but with the limitation that they neglect gravity, temperature, and plasma pressures and densities.

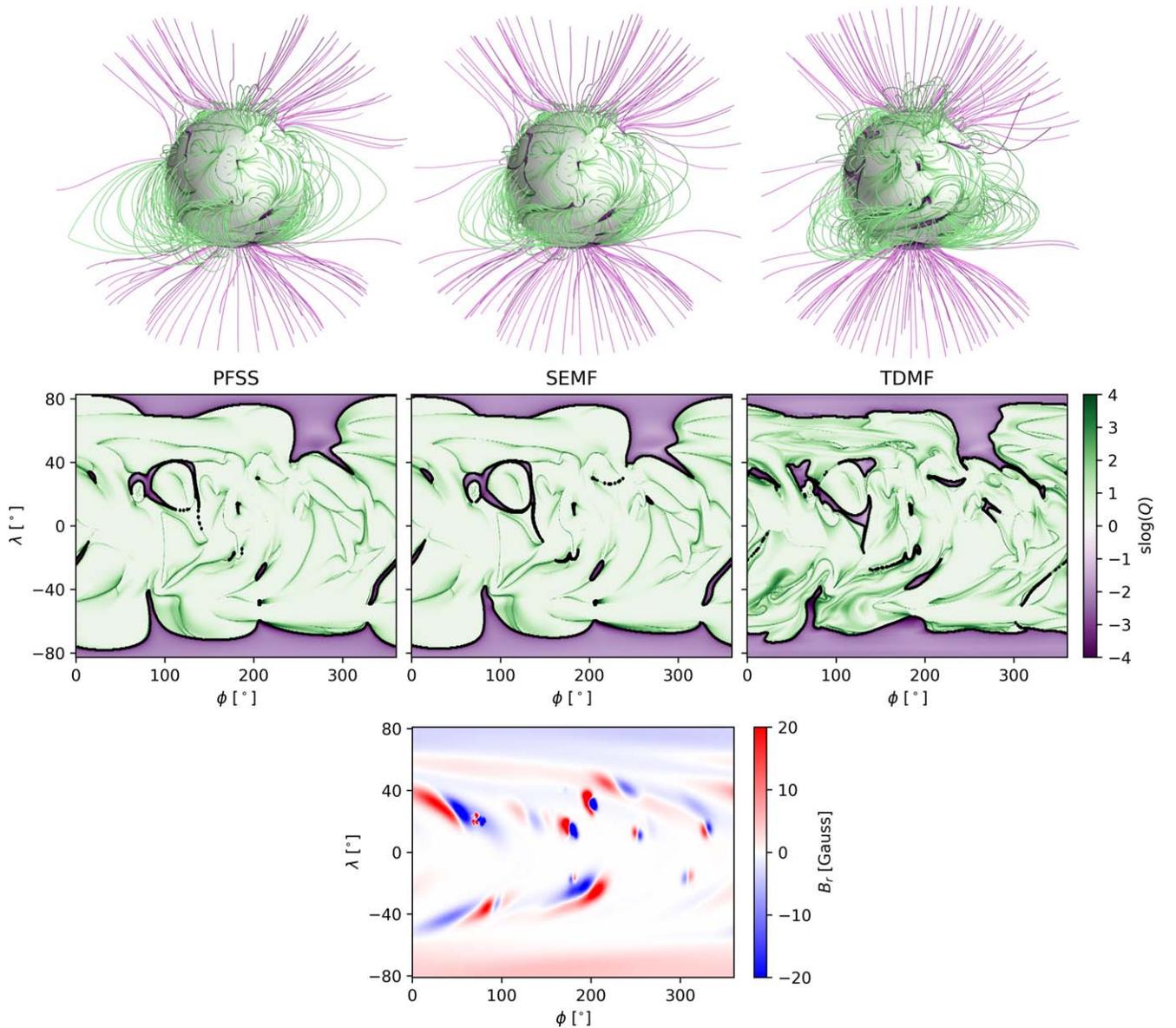

**Figure 1.** Top: a subset of field lines for the (left) PFSS model, (middle) SEMF model, and (right) TDMF model for simulations initialized with TDMF data from 2010 December 18. slog(Q) is plotted on the photosphere, and the plots are centered at a longitude of $\phi = 180°$. Middle: slog(Q) maps of latitude $\lambda$ against longitude $\phi$ overlaid with a black contour representing the OCB for each of the three models above. Bottom: the $B_r$ distribution on the photosphere for all three models (generated by the TDMF model).

Furthermore, the velocity for the magnetofrictional methods is simply an imposed function and uses an outflow profile that has only a radial dependence. It should therefore be noted that while the magnetofrictional methods are a practical alternative to the PFSS model, they are still a simplification compared to full MHD solutions and reality.

### 2.6. Mapping the Open–Closed Boundary

Toward understanding the role of interchange reconnection in the origin of the SSW, we need to characterize the OCB, where closed and open magnetic flux reconnect. After obtaining solutions for the coronal magnetic field at various dates using the three methods outlined above, we use the Universal Fieldline Tracer (UFiT) code (see V. Aslanyan et al. 2024) to trace magnetic field lines, determine their connectivity (i.e., open or closed), and calculate the squashing factor $Q$ (defined by V. S. Titov et al. 2002) within the domain. Although the calculation of $Q$ is not required to locate the OCB itself, this mathematical measure provides insight into the complexity and topology of the magnetic field. In particular, regions of high $Q$ are indicative of rapid changes in field-line connectivity, i.e., quasi-separatrix layers (QSLs), and have been linked to current-sheet formation and magnetic reconnection, which could play a role in shaping solar wind dynamics (D. I. Pontin & E. R. Priest 2022).

The signed logarithm of $Q$, denoted slog($Q$), is defined to be positive (negative) when the field is closed (open). The middle panels of Figure 1 show slog($Q$) on the photosphere for the three models, colored green (purple) to represent closed (open) field and overlaid with a black contour that represents the OCB. The location of the OCB is determined by the zero-contour of slog($Q$), where the magnetic field connectivity transitions between closed and open.

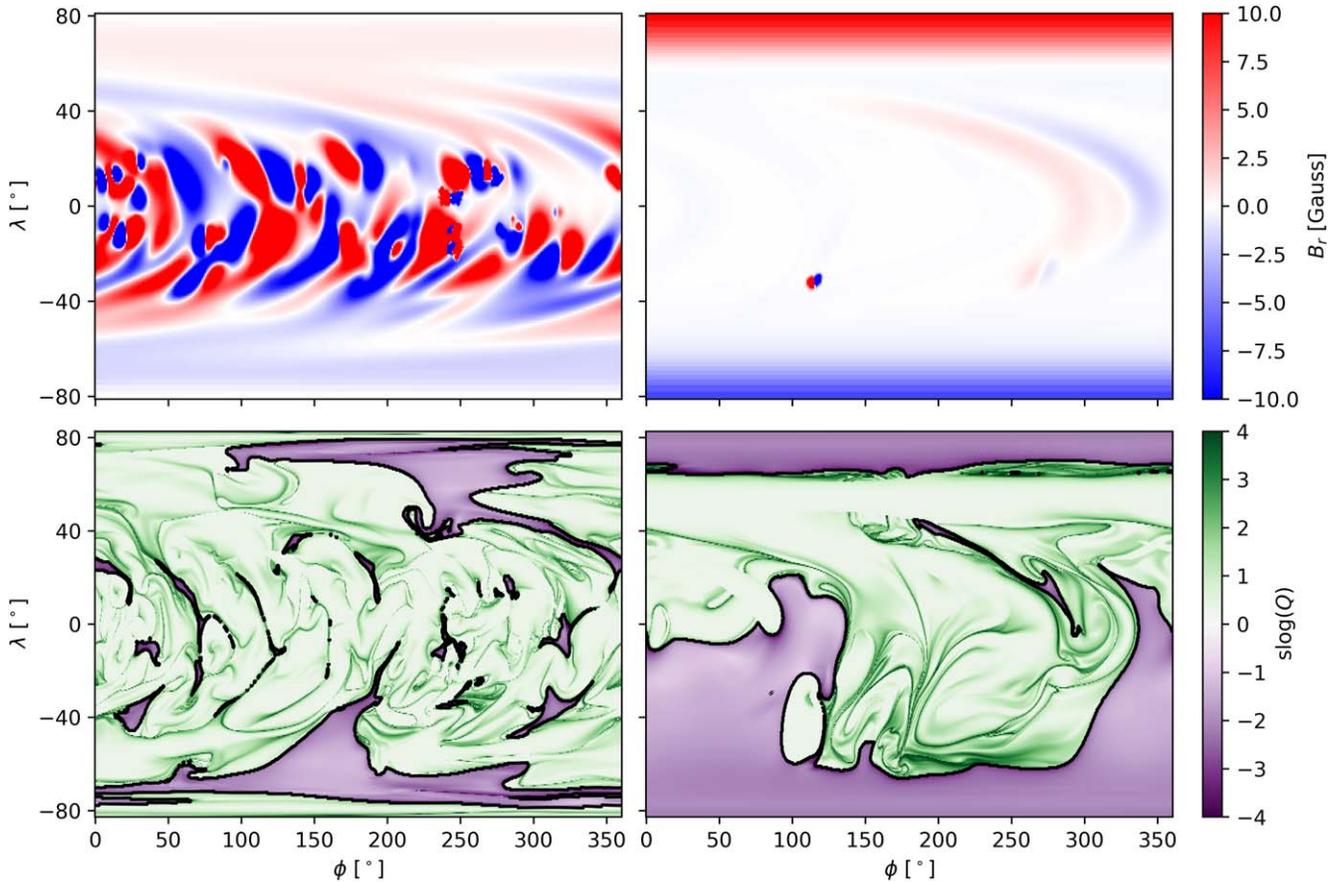

**Figure 2.** Top: $B_r$ distributions on the photosphere for the TDMF model around (left) solar maximum and (right) solar minimum (2014 April and 2019 December, respectively). Bottom: the corresponding slog($Q$) maps for the TDMF extrapolations. The black contour represents the OCB.

We evaluate $Q$ on a numerical grid of double the resolution of the underlying magnetic field grid, which ensures that for the chosen resolution of the boundary data, the OCB is well resolved (see Appendix A.2 for details). We confine ourselves to magnetogram resolutions of (180, 360) in ($\cos\theta, \phi$), and hence use a $Q$ computational mesh with resolution (360, 720) (for linearly spaced points in $\theta$ and $\phi$).

As expected, an input magnetogram with higher resolution or less smoothing will lead to an OCB that exhibits features on smaller length scales (see Appendix A.2). Therefore, the focus of our following analysis is the relative differences between models and throughout the solar cycle period, rather than absolute values of quantities (such as the OCB length), which can be influenced by factors like magnetogram resolution and smoothing.

### 2.7. Squashing Factor Complexity in the Time-dependent Magnetofrictional Model

Although a detailed analysis of $Q$ is beyond the scope of this paper, we note the high-$Q$ structures present in the closed field regions of each model in the middle panel of Figure 1. In all models, one or more QSLs are present within a given polarity as closed field lines connect to different neighboring polarities. In the TDMF model, this connectivity to neighboring polarities is more complex due to the time history, meaning that there are more, stronger high-$Q$ structures. Additionally, the TDMF model allows for large magnetic stresses to build in the lower corona. In closed field regions, this results in the formation of structures such as filament channels, which generally exhibit large $Q$ and contribute to the complexity of the field.

The overall complexity of the slog($Q$) maps changes over the solar cycle with the complexity of the photospheric magnetic field. This is demonstrated in Figure 2, which shows the $B_r$ magnetograms (top panel) and corresponding slog($Q$) maps (bottom panel) for the TDMF model at solar maximum (left) and solar minimum (right). At solar maximum, the ongoing flux evolution drives increasingly complex connectivity patterns near polarity inversion lines where opposing magnetic fluxes interact. At solar minimum, although the overall complexity is less, long high-$Q$ corridors remain prevalent in the TDMF model due to the gradual shearing of magnetic field lines as photospheric flux evolves. The persistence of these high-$Q$ features in the TDMF model across the solar cycle highlights the significance of the flux transport scheme in shaping the topology of the coronal magnetic field.

## 3. The Open–Closed Boundary over Solar Cycle 24

We begin our analysis in this section by examining the OCB itself, and focus on the structure throughout Solar Cycle 24 to characterize expected cycle variations. We then explore the nature of the flux in the vicinity of the OCB. We analyze the OCB on the lower boundary of our model (the photosphere); however, we note that dynamically important interchange reconnection—from the point of view of release of closed coronal plasma onto open field lines—is likely only the reconnection that occurs in the corona itself. From this point of view, it could be argued that analyzing the OCB at the base of the corona is more appropriate. However, the typical height for the transition region is within the first radial pixel in our models, and we have verified that the maximum relative difference in the OCB length between the surfaces $r = R_\odot$ and $r = R_\odot + 2.5$ Mm is 1%–2%. For simplicity, we therefore analyze the OCB on the photosphere only.

### 3.1. Characteristics and Model Dependence of Midlatitude Coronal Holes

The middle panel of Figure 1 shows the OCB structure from 2010 December 18 for the three coronal magnetic field models outlined in Section 2. As expected, because this is close to solar minimum, there is a large open region around each pole for each model, as well as localized open field regions at lower latitudes.

For the PFSS and SEMF models (left and middle panels), there is a similarity in the location of midlatitude CHs, which we define as those within $-45° < \lambda < 45°$ (where $\lambda$ is latitude). However, the CHs present in the PFSS model are often larger in the corresponding SEMF case, leading to larger open flux estimates. PFSS models, which rely on a fixed source-surface boundary to mimic the effects of the solar wind outflow, often underestimate the observed open magnetic flux. This underestimate can be mitigated by reducing the source-surface height, however this in turn causes CHs to expand (J. A. Linker et al. 2017) and prevents streamers from reaching higher altitudes, leading to a less realistic coronal magnetic field. The SEMF model overcomes this limitation by imposing an outflow wind, which leads to a more realistic coronal configuration that better accounts for some of this missing flux. For the results in Figure 1, we find that the midlatitude open flux for the PFSS model is ∼83% of that determined by the SEMF model.

The TDMF model (right panel of Figure 1) shows midlatitude CHs similarly located to those in the SEMF and PFSS models, but with notably different sizes and shapes. Many of the midlatitude open regions in the TDMF case are stretched over larger areas on the photosphere, so contain more open magnetic field than that of the previous two models. Namely, the midlatitude open flux for the SEMF case is ∼70% of that determined by the TDMF model, so there is a greater discrepancy here than between the PFSS and SEMF models. This suggests that the inclusion of the "memory" associated with the time evolution of the TDMF model can have a significant impact on determining the location and size of midlatitude open flux regions and hence the structure of the OCB. This is consistent with Z. Mikić et al. (2018), who found that a time-dependent model was crucial for reproducing eclipse observations.

### 3.2. Solar Cycle Dependence of the Open–Closed Boundary Length

Given that the three coronal models are initialized with the same input magnetogram and the midlatitude CHs appear similarly located between the models, the area occupied by these midlatitude CHs likely plays a significant role in the differences observed in the midlatitude open fluxes. For the results in Figure 1, we find that the area occupied by the midlatitude CHs for the PFSS model is ∼77% of that occupied by the SEMF model, and the area for the SEMF model is ∼65% of the TDMF model. Because midlatitude CHs are bounded by the OCB, we expect larger midlatitude open regions to correlate with longer OCB lengths.

The OCB length on the photosphere is calculated by summing the great circle distances between adjacent points on the OCB contours shown in the slog($Q$) maps for each model. As anticipated, the longest OCB is found in the TDMF model, followed by the SEMF and PFSS models. This is consistent with findings for the open flux and area occupied by the midlatitude CHs in each of these models.

This systematic offset in the OCB lengths is consistent across the solar cycle. Figure 3 shows the OCB length on the photosphere over the cycle period for the PFSS (cyan), SEMF (orange), and TDMF (purple) models. We also compare with a direct PFSS extrapolation of smoothed HMI synoptic maps from the `hmi.synoptic_mr_polfil_720s` series (see the black dashed curve in Figure 3). The details of this PFSS extrapolation are given in Appendix A.1.

Solar cycle trends of the OCB length on the photosphere are similar for the PFSS and SEMF models, since midlatitude CHs for these models have comparable shapes and are similarly located across the cycle period. Consequently, the differences in OCB lengths more likely arise from CHs in the SEMF model expanding into the closed field in the PFSS model as a result of the outflowing wind.

The TDMF model consistently reveals longer OCB lengths compared to the static models, with some different cyclic trends. Leading up to solar maximum, the TDMF model exhibits an overall increase in OCB length that is not observed in the other models. Examination of the slog($Q$) maps across different dates reveals that CHs in the TDMF model are not only larger and more abundant than in the other two models, but also significantly more complex in shape. This additional complexity in the boundaries of CHs in the TDMF model is likely the result of the more sheared and contorted magnetic field structures that build up in the coronal volume during the time evolution (see, e.g., the top-right panel of Figure 1). This results in more intricate and winding CH boundaries. This, coupled with the larger and more abundant midlatitude CHs in the TDMF model, ultimately yields longer overall OCB lengths relative to the static models.

In Figure 3, leading up to the solar maximum of Solar Cycle 24, the OCB lengths for the PFSS-extrapolated HMI data align better with the PFSS and SEMF models.[6] Following solar maximum, however, the PFSS-extrapolated HMI results show a closer agreement with the TDMF and SEMF models. A distinct jump in OCB length is observed at solar maximum in the PFSS-extrapolated HMI results, which is not present in the three models initialized with TDMF simulation data. This increase in the OCB length coincides with the opening of the southern polar CH during the pole reversal, which occurs at the very beginning of the substantial increase in the total open flux around solar maximum (S. G. Heinemann et al. 2024).

Finally, all four curves in Figure 3 show a gradual decline in OCB length following solar maximum and approaching the 2019 solar minimum. During this period, the midlatitude region becomes increasingly dominated by streamers compared to pseudo-streamers (i.e., closed field that separates regions of the same polarity). This occurs since fewer active regions are present during solar minimum, and the coronal magnetic field is evolving toward a dipole configuration (see Figure 2). By the 2019 solar minimum, the OCB predominantly consists of a contour enclosing the large polar CHs. Therefore, the decline in the OCB lengths following solar maximum can be attributed to the decrease in the abundance of midlatitude CHs and the overall topological changes in the coronal magnetic field.

---

[6] It should be noted that the length of the OCB is largely dependent on the chosen smoothing of the input HMI data, which we explore further in Appendix A.2.

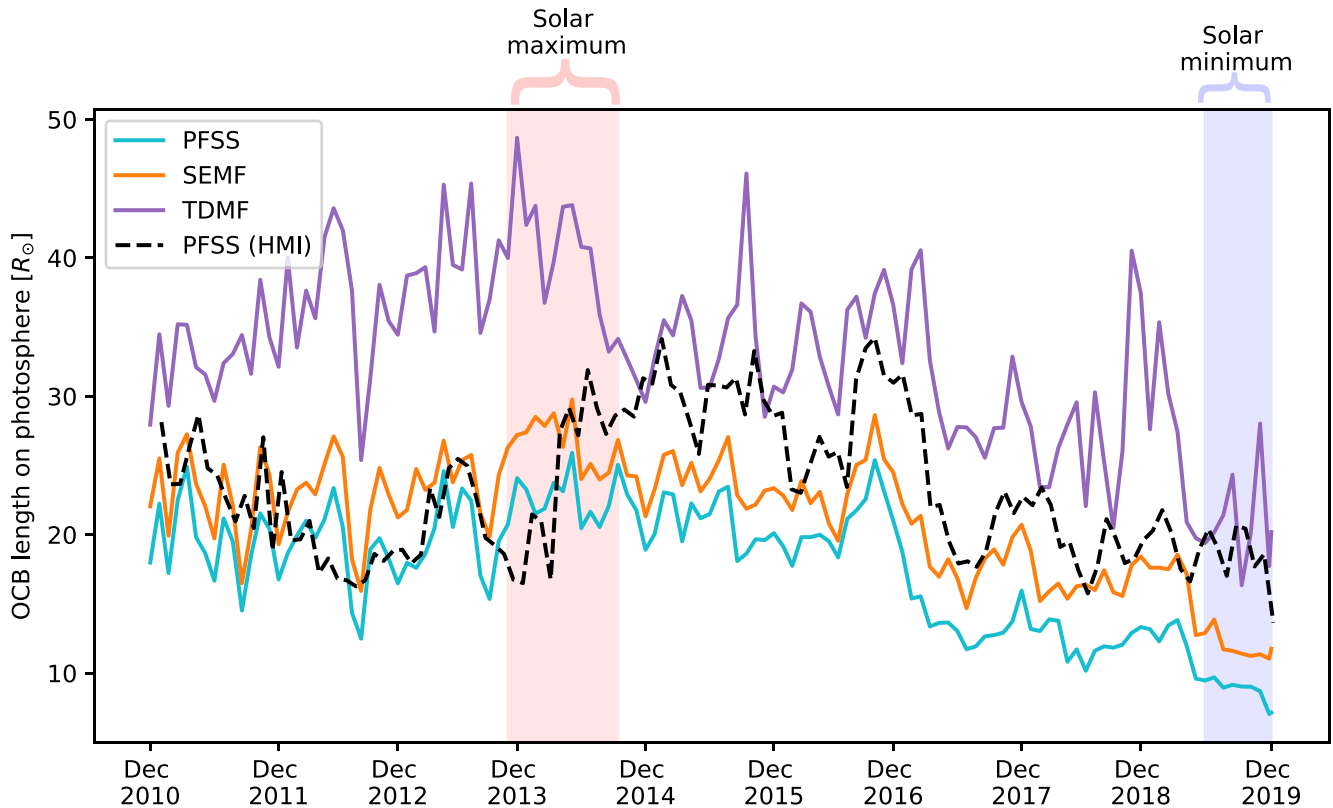

**Figure 3.** OCB length on photosphere (in units of solar radius, $R_\odot$) over Solar Cycle 24 for PFSS (cyan), SEMF (orange), TDMF (purple) models, and PFSS extrapolations of HMI data (black). The red shaded region indicates solar maximum, specifically ±6 months of 2014 April where the daily total sunspot number was at maximum (sourced from WDC-SILSO, Royal Observatory of Belgium, Brussels). The blue shaded region indicates solar minimum, specifically 6 months prior to 2019 December (where the sunspot number was at minimum).

## 4. The Magnetic Field Near the Open–Closed Boundary

Having examined the dependence of OCB length on both the model choice and the solar cycle phase, we now turn to characterizing the magnetic field on the photosphere in the vicinity of the OCB. In an idealized scenario where the Sun's global magnetic field is a pure dipole, S. K. Antiochos et al. (2011) suggest that dynamic broadening of the OCB at the photosphere—driven by supergranular flows and flux emergence (H. Schunker et al. 2024)—would occur at a scale consistent with supergranule sizes. C. N. Arge et al. (2024) and A. Koukras et al. (2023) further support this, identifying supergranules with a spatial scale of ∼25–30 Mm as the dominant driver of OCB dynamics. Motivated by these insights, we focus our analysis on the magnetic field within 1 supergranule diameter (i.e., 25 Mm onto each of the open and closed field sides) of the OCB. We refer to this as the *near-OCB* magnetic field.

In the sections below, we determine the near-OCB closed magnetic flux and compare our results with sunspot number measurements. We also investigate the length of the field lines in this region, as some studies suggest a link between coronal loop size and electron temperatures, which is also known to distinguish fast and slow wind (G. Gloeckler et al. 2003; U. Feldman et al. 2005). We then analyze the near-OCB open flux and calculate the area occupied by this flux on the source surface for comparison with IPS measurements of solar wind speeds.

### 4.1. Closed Magnetic Flux Near the Open–Closed Boundary

We calculate the near-OCB flux by identifying all flux that is rooted at the photosphere within a fixed distance of the OCB, chosen to be 25 Mm. To find this flux, we identify every pixel on the photosphere whose center lies with 25 Mm of the OCB; this then forms a "corridor," 5–10 pixels wide depending on the latitude, around the OCB.

Figure 4 shows some properties of the near-OCB closed field for the TDMF model for 2010 December 18. The top-left panel shows the $B_r$ distribution on the photosphere, overlaid with the OCB in black. The bottom-left panel shows the distribution of unsigned flux in the near-OCB closed field. A high concentration of flux is observed near the edges of midlatitude CHs due to the strong $B_r$ of active regions in the corresponding magnetogram, which are clearly colocated with the midlatitude CHs. This suggests that there may be some correlation between the net unsigned flux in the near-OCB closed field and sunspot numbers recorded across Solar Cycle 24.

The top panel of Figure 5 shows the total unsigned flux in the near-OCB closed field for the four coronal field extrapolation methods over time. The tendency for the flux to be greatest in the TDMF followed by the SEMF and PFSS models is expected given the trends in the OCB lengths in Figure 3. A larger abundance of midlatitude CHs results in more near-OCB closed pixels, and consequently a larger total unsigned flux. All four methods show a peak in the total unsigned flux around solar maximum, followed by a decline toward the 2019 solar minimum. This trend closely matches the variations in the monthly mean sunspot numbers, shown in green in the bottom panel of Figure 5. The correlation is close in the case of the TDMF model (shown in purple).

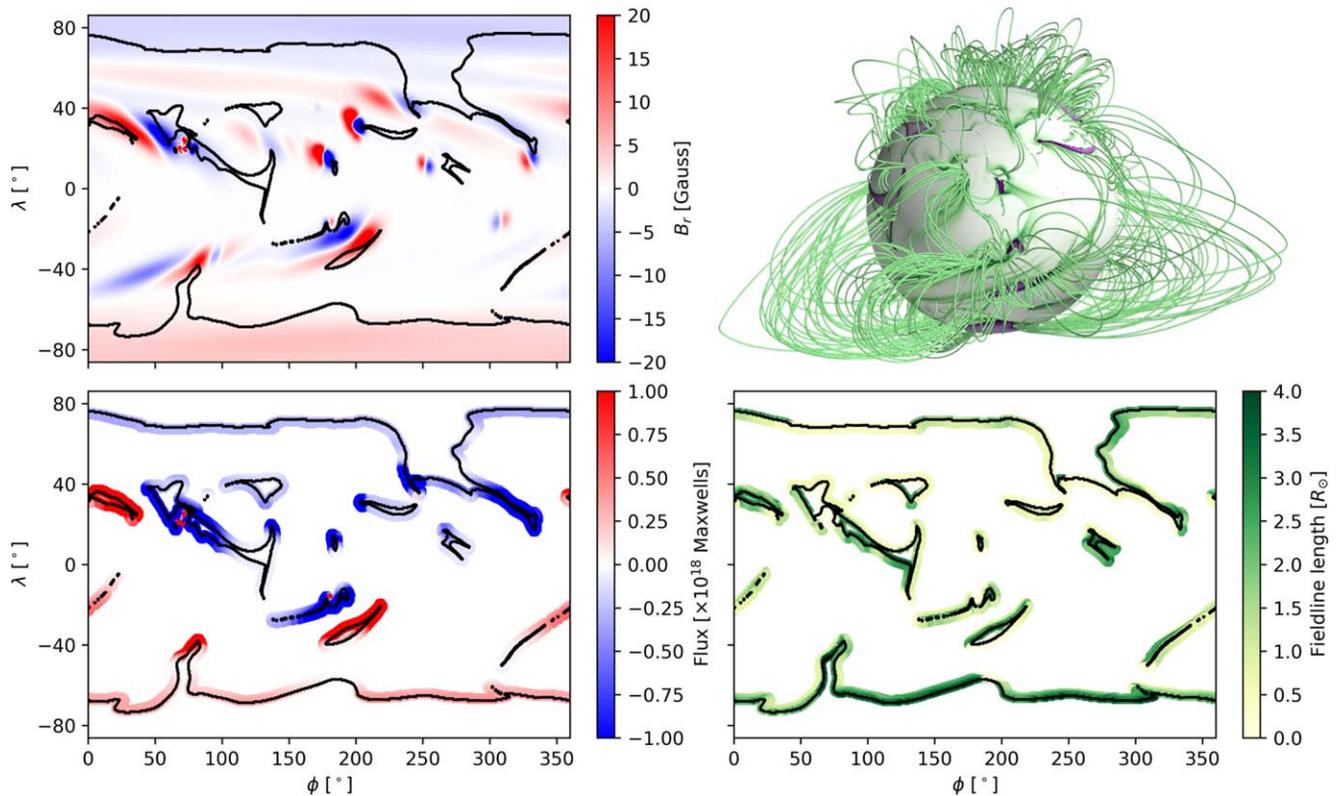

**Figure 4.** Results for the near-OCB closed field for TDMF simulations from 2010 December 18. Top: (left) the $B_r$ distribution for this date, overlaid with the OCB contour in black and (right) field-line tracings of a selection of near-OCB closed field lines. Bottom: near-OCB closed field pixels colored according to (left) signed flux-per-pixel in Maxwells and (right) field-line lengths in units of solar radius, $R_\odot$.

The top-right panel of Figure 4 shows a subset of the field lines traced from each near-OCB closed pixel for the TDMF model. The bottom-right panel shows the distribution of lengths of each near-OCB closed field line.[7] For the midlatitude CHs, a variety of field-line lengths are evident depending on their connectivity, e.g., field lines that connect opposite polarities of the same active region, neighboring active regions, or midlatitude active regions to opposite-polarity CHs at the poles. The length also depends on the proximity of footpoints to the OCB: Field lines closer to the OCB are longer as they are bounded by open field, forcing them to extend farther toward the top boundary than those that are rooted in closed field away from the OCB.

The predominance of long versus short field lines may also correlate with active region trends over the solar cycle. For example, near solar minimum, fewer active regions are observed (see Figure 2) and we therefore expect to see fewer pseudo-streamers in the midlatitude regions. As the coronal magnetic field approaches a dipole configuration, we expect the field to mostly consist of longer field lines connecting the opposite-polarity polar CHs. Figure 6 shows the percentage of the near-OCB closed flux attributed to field lines of various lengths. (It should be noted that this percentage did not show much variation between the models, and hence we only present the results for the TDMF model here.) These results are binned according to various length ranges to capture changes in the ratio of long to short streamers over the solar cycle. Interestingly, long field lines (i.e., those $>1.0 R_\odot$ in length) still contribute around 50% of the flux around solar maximum. As expected, we see an increase in the flux associated with longer field lines (to about 80%) around solar minimum. This is consistent with the less dynamic coronal structure and the dominance of global-scale magnetic field features during this phase of the solar cycle.

---

[7] The lengths were calculated by summing the Euclidean distance between adjacent points along the field lines traced from each near-OCB closed pixel.

### 4.2. Open Magnetic Flux Near the Open–Closed Boundary

Having explored the closed magnetic flux in a 25 Mm vicinity of the OCB, we now turn to the flux in the near-OCB open magnetic field, along which interchange-released plasma in the SSW may flow out into the heliosphere.

Figure 7 shows the solar cycle variation of the open flux for the different coronal models: PFSS (cyan), SEMF (orange), TDMF (purple), and PFSS extrapolations of HMI data (black). The top panel shows the total unsigned open magnetic flux over the entire solar surface. The SEMF model consistently predicts higher open flux than the PFSS model due to the radial outflowing wind that forces field lines near CH edges to open. The further enhancement of the open flux in the TDMF model results from the additional energization of the low corona, an effect also noted by A. R. Yeates et al. (2010), who demonstrated that the emergence of twisted active regions and large-scale shearing by photospheric motions increases the open flux compared to PFSS solutions. By contrast, PFSS-derived open flux strongly depends on the photospheric magnetogram input, accounting for the differences observed between the PFSS extrapolations of the HMI versus TDMF data (noting also that the HMI results are dependent on the smoothing of the input magnetogram). The jump in the open flux observed after solar maximum in all three models has been explored in detail in papers such as S. G. Heinemann et al. (2024), who suggest that this is linked to the emergence of a large bipole that caused the southern polar CH to expand and additional flux to open.

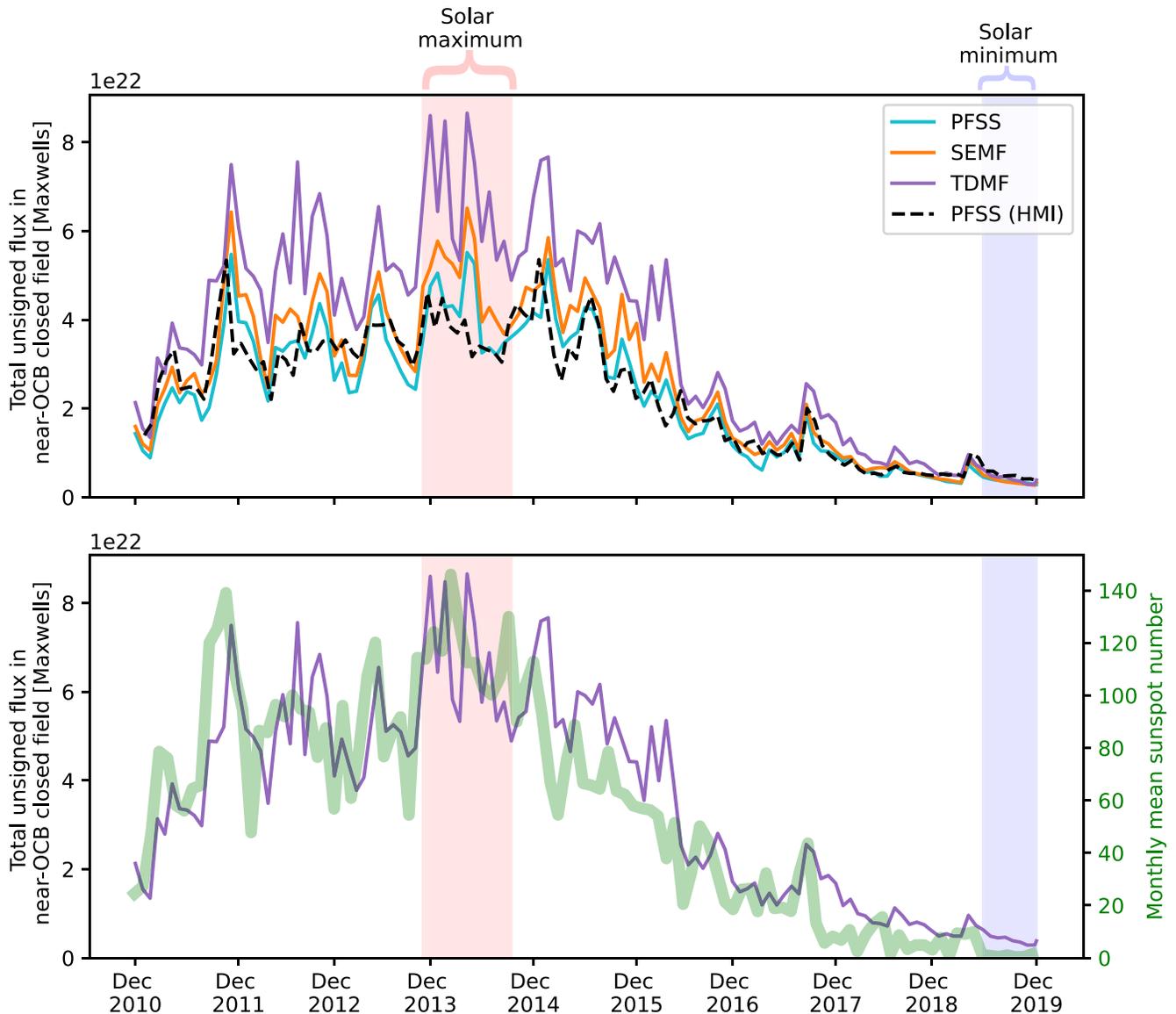

**Figure 5.** Top: net unsigned flux on the photosphere in near-OCB closed field for PFSS (cyan), SEMF (orange), TDMF (purple) models, and PFSS extrapolations of HMI data (black). Bottom: TDMF results (purple) plotted with monthly mean sunspot number (green), sourced from WDC-SILSO, Royal Observatory of Belgium, Brussels. The red (blue) shaded regions indicate solar maximum (minimum).

The second panel of Figure 7 shows the net unsigned flux in the near-OCB open field on the photosphere. As with the near-OCB closed flux in Figure 5, the largest estimate is consistently given by the TDMF model, and all four models exhibit a steady decline in the flux after solar maximum. The differences in the near-OCB open flux between the models are a direct result of the OCB length difference associated with CHs at lower latitudes in the vicinity of active regions, as discussed previously.

The third panel of Figure 7 shows the ratio of the second panel to the first—that is, the percentage of the total open flux attributed to the near-OCB open field on the photosphere. Very high percentages (between 80% and 100%) are evident leading up to solar maximum as the polar CHs decrease in size (see Figure 2) and the coronal field becomes dominated by streamers and pseudo-streamers. Consequently, the near-OCB open flux makes up a large fraction of the total open flux (as the width of the midlatitude CHs becomes comparable to the width of the near-OCB band). The relative contribution of the near-OCB open flux decreases toward solar minimum due to the expansion of the polar CHs and the dominance of open flux associated with the dipolar field.

Surprisingly, the percentages of near-OCB open flux agree closely across the three coronal field extrapolation methods, despite the model variations evident in the top two panels of Figure 7. A possible explanation lies in the scaling behavior of flux with CH size: For an idealized circular CH of radius $\mathcal{R}$ with uniformly distributed flux, the total open flux, $F_O$, scales with the CH area ($\propto \mathcal{R}^2$), while the near-OCB open flux, $F_{\rm O,OCB}$, scales approximately with the perimeter ($\propto \mathcal{R}$). In this scenario, variations in the CH size produce larger relative changes in $F_{\rm O}$ than in $F_{\rm O,OCB}/F_{\rm O}$. Indeed, we find this to hold when we compare a few instances of colocated CHs between the PFSS and SEMF models. Since both models are initialized with the same input magnetogram, this suggests that the CHs present in the PFSS model largely retain their structure in the SEMF model while expanding under the influence of the outflowing wind. At the other extreme of small, low-latitude CHs, often all of the flux lies within 25 Mm of the OCB, and so the fraction $F_{\rm O,OCB}/F_{\rm O} \sim 1$ again changes slowly relative to $F_{\rm O}$. In the TDMF model, midlatitude CHs tend to be more numerous and larger, but also longer and narrower. Averaged across all CHs, there is a proportional increase in both $F_{\rm O}$ and $F_{\rm O,OCB}$, preserving the fraction $F_{\rm O,OCB}/F_{\rm O}$ derived in the static models. Significantly, the close agreement of the curves in the third panel of Figure 7 implies that the fraction of near-OCB open flux is robust to the assumptions of the different coronal magnetic field models.

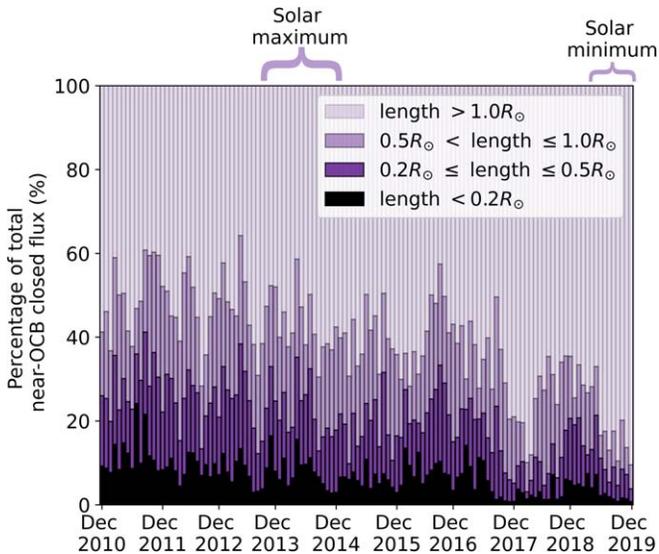

**Figure 6.** Percentage of flux in near-OCB closed field for the TDMF model, binned according to field-line lengths: less than $0.2R_\odot$, between $0.2R_\odot$ and $0.5R_\odot$, between $0.5R_\odot$ and $1.0R_\odot$, and greater than $1.0R_\odot$.

In the bottom panel of Figure 7, we show the percentage of the total area occupied by the near-OCB open field on the outer boundary ($r = 2.5R_\odot$). This was determined by tracing open magnetic field lines "inwards" from the outer boundary (from a grid of (360, 720) points equally spaced in ($\theta$, $\phi$)), and summing the area of the pixels at $r = 2.5R_\odot$ for which the open field line landed within 25 Mm of the OCB on the photosphere. Relevant observational data here includes IPS measurements of the solar wind speed distribution over Solar Cycle 24, detailed in M. Tokumaru et al. (2010) and M. Tokumaru et al. (2021). Specifically, M. Tokumaru et al. (2021, their Figure 15) shows the fraction of the source-surface area (where the IPS observations were made, at $2.5R_\odot$) occupied by wind of different speed ranges during the period 1985–2019. For Solar Cycle 24, M. Tokumaru et al. (2021) find that the fraction (by area) of slow wind (which we take to be their bins of less than 530 km s$^{-1}$) peaks at ~80% in 2011 and drops steadily to ~40% in 2019. Referring to the bottom panel of Figure 7, we find that in each of our models the percentage of the source-surface area occupied by the near-OCB open flux is largest leading up to and during solar maximum (where it sits between ~70% and 100%) and declines following the maximum to ~40% during solar minimum. This correlation between our results and the IPS measurements supports the hypothesis that SSW is transported into the heliosphere by open field lines that originate near the OCB.

## 5. Conclusions

In this study, we explore the global structure of the Sun's coronal magnetic field and the magnetic flux in the vicinity of the OCB. Such analysis is necessary for advancing our understanding of the heliosphere, since the flux within a narrow band around the OCB (of width 1 supergranule diameter, or 25 Mm) is thought to be essential for explaining the observed properties of the SSW (e.g., A. Koukras et al. 2023; C. N. Arge et al. 2024). As discussed in Section 4.2, the area-fraction of the total open flux that originates within this band hovers around 90% leading into solar maximum, and drops to less than 40% around solar minimum. This fraction aligns closely with IPS measurements of the solar wind over Solar Cycle 24 (M. Tokumaru et al. 2021, their Figure 15), which showed that ~80% of the observed wind was slow around solar maximum compared to ~40% at solar minimum. This implies that the calculation from the models of the fraction of near-OCB open flux could be used for predicting the amount of SSW in the heliosphere during different phases of the solar cycle.

The result that, near maximum, close to 100% of the open flux lies within 1 supergranule of the OCB is by itself an important finding. Note that the amount of open flux during this time of the cycle generally exceeds that near minimum. Our findings imply, therefore, that the CHs during maximum are primarily very narrow and stretched out, and include the strong field of active regions. Consequently, the solar wind, heliospheric magnetic field, and plasma during maximum are likely to have significantly different properties than during minimum. Our results would imply, for example, that the occurrence of structures such as switchbacks (J. C. Kasper et al. 2019), which are believed to be due to interchange reconnection, may be more common near maximum.

Furthermore, examining the fraction of the total open flux that lies within 25 Mm of the OCB on the photosphere, we find that it is independent of underlying model assumptions. In other words, the inclusion of electric currents and the dynamic buildup of magnetic stress in the corona appears not to impact this fraction. Significantly, this means that we can predict the expected totals of slow and fast wind using a simple and computationally efficient PFSS model.

Although the open flux fraction appears to be independent of the magnetic field dynamics, the detailed open field structure is not. From Figure 1, we note that the two static models (PFSS and SEMF) yield similar-looking CHs, but the dynamic model (TDMF) yields noticeably different open field structure. This is apparent also from the length of the OCB in Figure 3. In principle, the TDMF model should be more accurate, because it includes more of the dynamics known to be present on the real Sun. However, the TDMF model does not include the small-scale supergranular motions believed to be responsible for the release of the solar wind. It may be that when these are included, the resulting interchange reconnection is so frequent that the time-averaged system is closer to the steady models. The situation for the OCB may be similar to what has been proposed for the closed corona, in that the presence of ubiquitous and

                                                                                                                                              Wilkins et al.

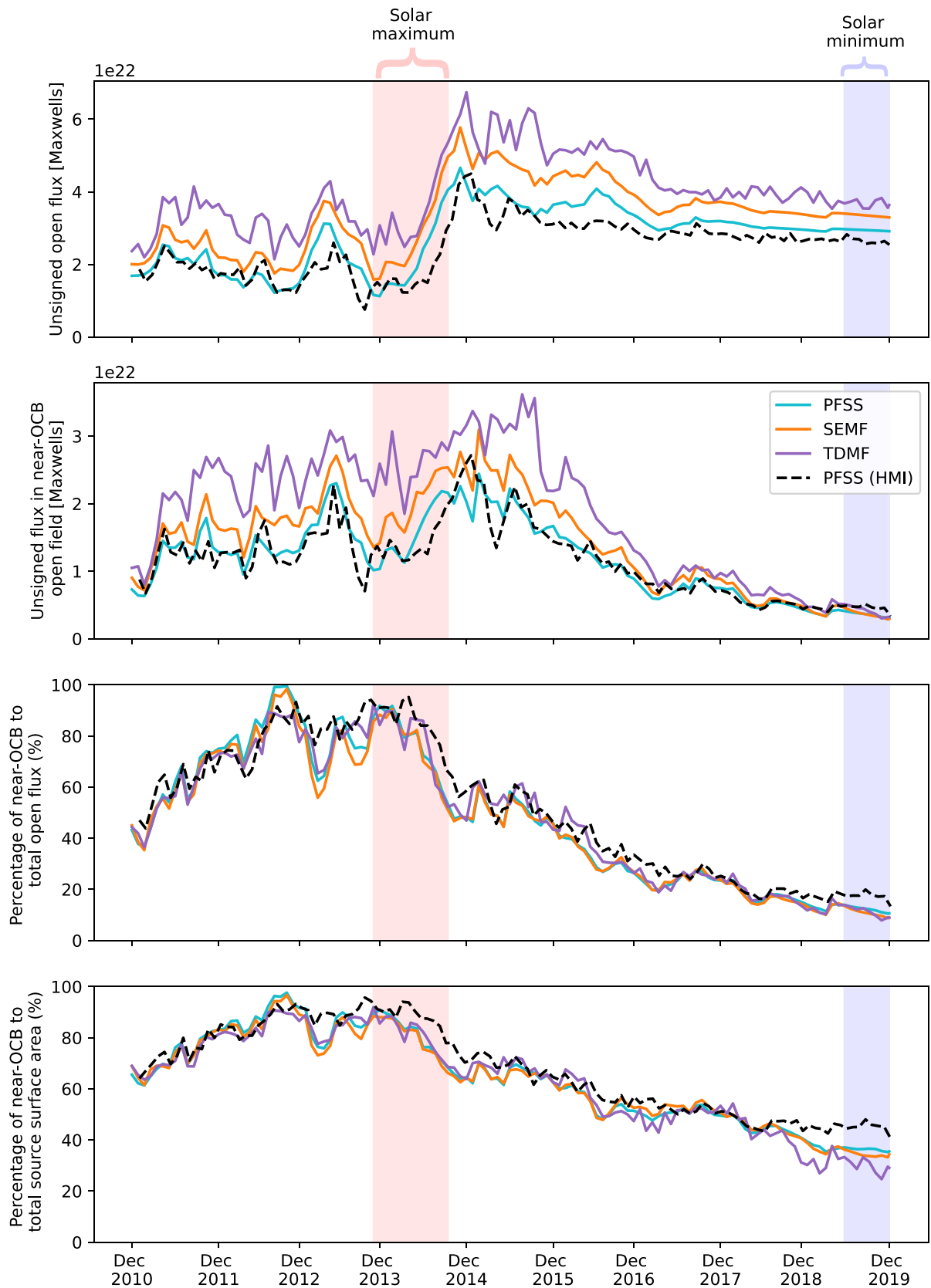

**Figure 7.** Open flux calculations for PFSS (cyan), SEMF (orange), TDMF (purple) models, and PFSS extrapolations of HMI data (black). Top: total unsigned open flux over the entire solar surface. Second: total unsigned near-OCB open flux on the photosphere. Third: the percentage of near-OCB to total open flux on the photosphere. Bottom: the percentage of area occupied by near-OCB open field at $2.5R_\odot$. The red (blue) shaded regions indicate solar maximum (minimum).

constant small-scale reconnection leads to a quasi-potential magnetic field (S. K. Antiochos 2013). More theoretical and observational work is needed to resolve this intriguing issue.

The overall length of the OCB on the photosphere may be important for understanding the net transfer of mass and momentum between the open and closed field. We find that this length is more sensitive to the underlying model than the solar cycle phase. A longer OCB results when the field is nonpotential and an outflow velocity is present. This length increases further when a driven photospheric boundary is included and the sheared magnetic field in the corona is able to evolve over time (i.e., in the TDMF model). As a result of this time dependence, midlatitude CHs are more abundant (particularly around solar maximum), and undergo changes in their shape and complexity. This also leads to a systematic increase in the total open flux compared to the static models. However, no correlation is observed between the total open flux and OCB length over the solar cycle. This suggests that the total open flux is less dependent on CH size and more on the proximity of those CHs to regions of strong $B_r$. This is evidenced by the fact that the open flux is consistently greatest for the TDMF model where the CHs are most strongly correlated with active region locations.

The height in the corona at which interchange reconnection occurs likely affects the charge-state ratios within the released plasma. As such, we analyze the length of the closed field lines in the vicinity of the OCB where this reconnection takes place. We find that the flux contribution of shorter ($<0.5R_\odot$) field lines is larger during solar maximum than minimum. This may account for the well-known variation of solar wind plasma elemental and charge-state composition during the solar cycle (T. H. Zurbuchen et al. 2002). Moreover, the near-OCB closed flux follows sunspot number trends, peaking around solar maximum and decreasing toward solar minimum. These results reinforce the hypothesis that midlatitude CHs, particularly those colocated with active regions, contribute more to the near-OCB flux during solar maximum.

Our findings regarding the solar cycle and model dependence of OCB characteristics reinforce the need for further investigation of near-OCB dynamics and how this relates to interchange reconnection and the origin of the SSW. Future work should include analysis of the topology of high-$Q$ structures in slog($Q$) maps at different heights and over the solar cycle. Calculations of the expansion factor—an important indicator of fast and slow wind sources—may also provide insight into the relationship between near-OCB open flux and the SSW. Additionally, the link between charge-state ratio measurements of the solar wind and coronal loop lengths and heights needs to be determined. Such analyses, and comparisons with observations especially, will help to clarify the mechanisms governing open–closed flux interactions and advance our understanding of SSW sources throughout the solar cycle.

## Acknowledgments

D.P. gratefully acknowledges support through an Australian Research Council Discovery Project (grant No. DP210100709). H.S. is the recipient of an Australian Research Council Future Fellowship Award (project number FT220100330). This research is funded by a grant from the Australian Government. S.K.A. acknowledges support for this research from the NASA LWS and NSF SHINE Programs. C.W., D.P., and H.S. acknowledge the Awabakal people, the traditional custodians of the unceded land on which their research was undertaken at the University of Newcastle.

# Appendix
# PFSS Extrapolations of Different Input Magnetograms

This appendix compares PFSS extrapolations initialized with two distinct data sets: the TDMF simulation results and the smoothed HMI synoptic maps. For the latter case, we analyze how variations in the smoothing of the input magnetogram and the computational grid resolution influence the OCB structure at different heights above the photosphere.

## A.1. Initializing the Model with TDMF versus HMI Magnetograms

For various figures in this paper, we compare the results for the three coronal magnetic field models initialized with the TDMF data with a direct PFSS extrapolation of (smoothed) HMI data from the hmi.synoptic_mr_polfil_720s series. We use data from Carrington rotations CR2105 to CR2225 (which correspond approximately to the TDMF simulation dates). The HMI magnetograms are smoothed with a pseudo-Gaussian filter $\exp(-l(l+1)k)$ for spherical harmonic $l$ with $k = 10^{-4}$. The smoothing parameter $k$ is chosen such that the total flux on the photosphere for the PFSS extrapolations of the HMI data approximately matches that of the PFSS extrapolations of the TDMF simulation data across the solar cycle. The HMI magnetograms were resampled to the same resolution as the TDMF input magnetograms, i.e., (180, 360) in ($\cos\theta$, $\phi$), and the squashing factor calculations were performed on a computational grid with resolution (360, 720) in ($\theta$, $\phi$).

Figure 8 compares $B_r$ on the photosphere for the first simulation date in the TDMF and HMI series (top panel) and shows the corresponding slog($Q$) maps and OCB for the PFSS extrapolations of these magnetograms (bottom panel). Some notable differences in the active region structures are evident, driven by both the intrinsic properties of the data sets and the smoothing. This in turn influences the connectivity of the magnetic field, including the location of open flux regions and the OCB. We demonstrate in the next subsection that these features are influenced significantly by the smoothing of the input magnetogram.

## A.2. The Effect of Smoothing the Input Magnetogram

To investigate the impact of magnetogram smoothing, we examine the $B_r$ magnetograms, slog($Q$) maps, and OCB contour for different values of the smoothing parameter $k$ in the pseudo-Gaussian filter applied to the HMI magnetogram for CR2105. These results are presented in Figure 9.

As expected, less smoothing results in more complex $Q$ structures, characterized by finer-scale topological features. Consequently, additional open field regions emerge at locations that are closed when a larger smoothing $k$ is used, and more small-scale OCB contours are evident. This highlights how the choice of smoothing significantly influences the proportion of open and closed flux and the overall length of the OCB on the photosphere.

In Figure 10, we examine how the OCB length varies with height above the photosphere, focusing on the effects of the smoothing parameter $k$ and the computational $Q$ grid resolution. In the left panel, the $Q$ grid resolution is fixed at (360, 720) in ($\theta$, $\phi$), while the smoothing parameter varies between $k = 5 \times 10^{-5}$ (blue), $k = 10^{-4}$ (black), $k = 5 \times 10^{-4}$ (pink), and $k = 10^{-3}$ (yellow). In the right panel, the smoothing parameter is fixed to $k = 10^{-4}$, and $Q$ grid resolutions that are the same as (blue), double (black), and triple (pink) the resolution of the input magnetogram are explored.[8]

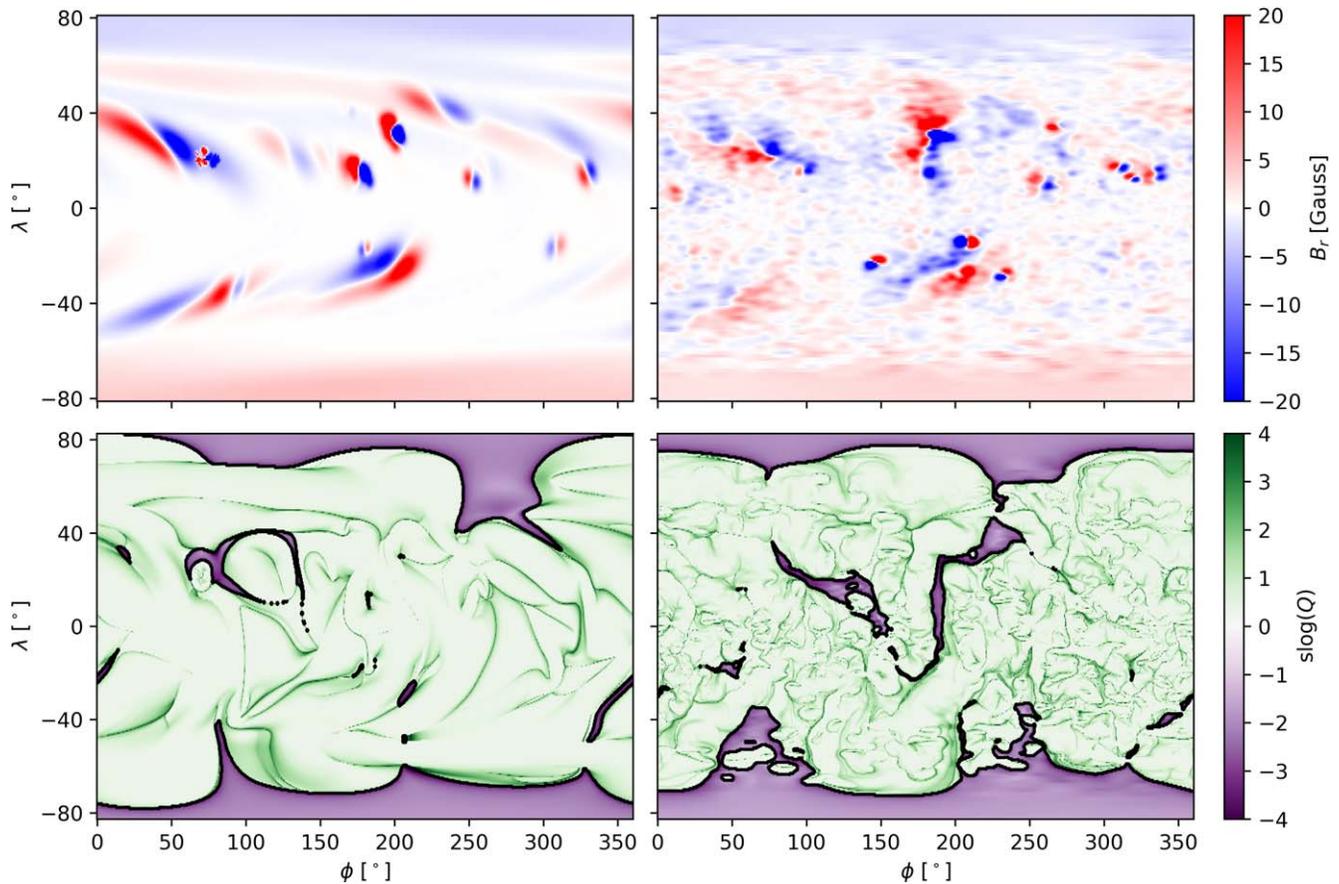

**Figure 8.** Top: $B_r$ magnetograms for the first simulation date in the (left) TDMF and (right) HMI series. These are 2010 December 18 and CR2105 (around 2010 December to 2011 January), respectively. Bottom: slog($Q$) maps on photosphere for PFSS extrapolations of the above magnetogram inputs, overlaid with the OCB in black.

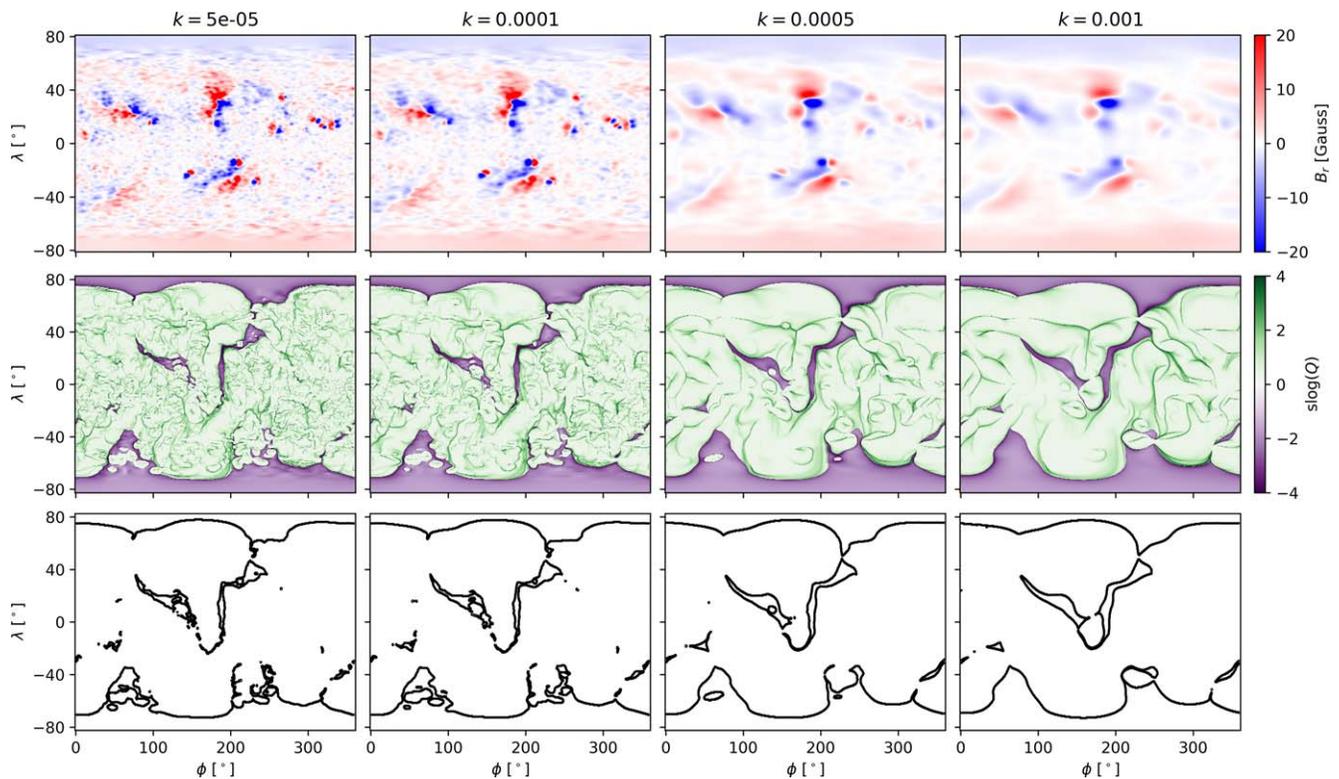

**Figure 9.** Effect of the smoothing parameter $k$ in the pseudo-Gaussian filter applied to HMI data for CR2105. From left to right, results for the values $k = 5 \times 10^{-5}$, $10^{-4}$, $5 \times 10^{-4}$, and $10^{-3}$ are shown. Top: $B_r$ magnetograms after smoothing and resampling. Middle: slog($Q$) on the photosphere for the corresponding PFSS extrapolations, colored according to open (purple) or closed (green) field. Bottom: the OCB contour generated from the slog($Q$) results.

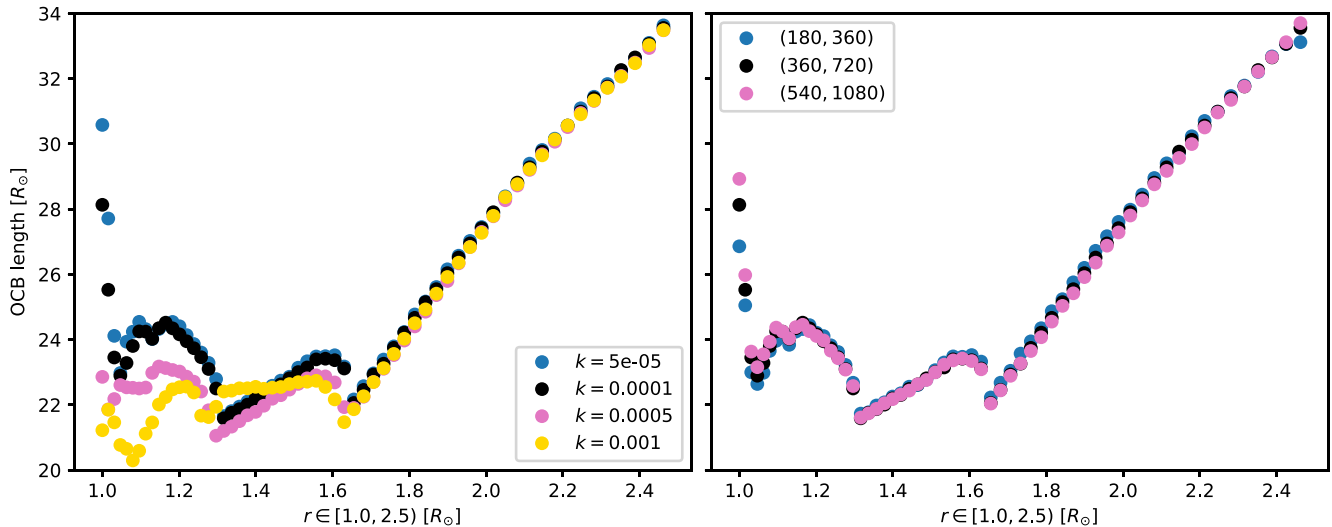

**Figure 10.** OCB length as a function of height for various magnetogram smoothing and $Q$ grid resolutions for PFSS-extrapolated HMI data for CR2105. Left panel: OCB lengths for smoothing parameter values of $k = 5 \times 10^{-5}$ (blue), $k = 10^{-4}$ (black), $k = 5 \times 10^{-4}$ (pink), and $k = 10^{-3}$ (yellow). Right panel: OCB lengths for various $Q$ computational grid resolutions (where the input magnetogram has fixed smoothing $k = 10^{-4}$ and resolution (180, 360)): (180, 360) (blue), (360, 720) (black), and (540, 1080) (pink) in $(\theta, \phi)$.

As expected, the OCB length converges at larger heights due to radial expansion. At lower heights (above the photosphere), the OCB lengths converge as the smoothing is decreased. The OCB length shows minimal sensitivity to changes in the $Q$ grid resolution for a fixed magnetogram resolution, as per the right panel of Figure 10. Increasing the grid resolution smooths existing OCB contours but does not greatly impact the topology of the slog($Q$) maps and OCB, which depend more on factors like the input magnetogram resolution and smoothing.

Lastly, we note that in this paper we have chosen to examine the OCB only on the photosphere for ease of exposition, however our method can be applied to any surface. As shown in Figure 10, the OCB length grows approximately linearly with radius above $\sim 1.6 R_\odot$ due to radial expansion. At lower heights, particularly near solar maximum, the OCB is more complex. As shown in Figure 1—and analyzed in much more detail in R. B. Scott et al. (2018, 2019)—the closed field region "bulges" into the open field region in a number of places, forming quasi-circular (partially) disconnected structures characteristic of separatrix domes that enclose strong field concentrations. As we go up in altitude, a larger number of disconnected closed regions can appear (as the chosen surface cuts through the apices of these domes). Looking at the OCB length as a function of height, we often observe one (or more) local maxima occur as a result—typically around 1.1–1.6$R_\odot$—as shown in Figure 10.

## ORCID iDs

Chloe P. Wilkins ⓘ https://orcid.org/0000-0002-0154-8380
David I. Pontin ⓘ https://orcid.org/0000-0002-1089-9270
Anthony R. Yeates ⓘ https://orcid.org/0000-0002-2728-4053
Spiro K. Antiochos ⓘ https://orcid.org/0000-0003-0176-4312
Hannah Schunker ⓘ https://orcid.org/0000-0001-9932-9559

---

[8] Note that the results presented in this paper use smoothing $k = 10^{-4}$ for the HMI magnetograms and a fixed $Q$ grid resolution that is double that of the input magnetogram.